\documentclass[%
 aip,
apl,%
 amsmath,amssymb,
 reprint,%
]{revtex4-1}

\usepackage{epsfig}
\usepackage{dcolumn}
\usepackage{bm}

\begin{document}


\title[Sample title]{Comparing the distribution of the electronic gap\\ of an organic molecule with its photoluminescence spectrum}%

\author{D.C. Milan}
\author{J.M. Villalvilla}
\author{M.A. D\'\i az-Garc\'\i a}
\author{C. Untiedt}
 \email{untiedt@ua.es}
 
 \affiliation{ 
Departamento de F\'\i sica Aplicada and Instituto Universitario de Materiales de Alicante.\\ Facultad de Ciencias (Fase II). Universidad de Alicante. E-03080 Alicante. Spain
}

\date{\today}

\begin{abstract}
The electronic gap structure the organic molecule N,N'-diphenyl-N,N'-bis(3-methylphenyl)-(1,1'-biphenyl)-4,4'-diamine, or TPD, has been studied by means of a Scanning Tunneling Microscope (STM) operated under ambient conditions, and by Photoluminescence (PL) analysis. 
Thousands of current-voltage characteristics have been measured at different spots of the sample showing the typical behavior of a semiconductor. The analysis of the curves allows us to construct  a gap distribution histogram which reassembles the PL spectrum of this compound. Our analysis shows that in the TPD films most of the observed distribution of the gap comes mainly from an uncertainty in the position of the LUMO levels of the molecular compound as would be expected from theory. This analysis demonstrates that STM can give relevant information, not only related to the expected value of a semiconductor gap, but also on its distribution which affects its physical properties such as the case of the PL and absorption distributions as here is reported. 
\end{abstract}

\pacs{73.61.Ph, 81.05.Fb, 85.65.+h, 33.20.-t, 85.60.Bt, }
\keywords{STM, Organic Molecules, Semiconductor Gap, Photoluminescence }

\maketitle

Local variations of the Density of States of a material can give rise to small changes of its physical properties which turn into a macroscopical uncertainty when averaged.
This study drove Binning and Rohrer to develop  a new tool to make electronic spectroscopy at the local scale which became the basis of a new microscope, the Scanning Tunneling Microscope (STM).\cite{Binning_82PRL,Binning_82JAP} 
Different spatially-resolved spectroscopic methods were implemented to this technique, generically known as Scanning Tunneling Spectroscopy (STS), \cite{Stroscio_PRL86,Feenstra_94}
through the analysis of the differential conductivity, providing the possibility of measuring the surface local density of states (LDOS).
Over the two last decades, different groups have studied through STS different materials to characterize the electronic properties of surfaces and adsorbed molecules starting from the early measurements on semiconducting surfaces\cite{Stroscio_PRL86} or molecular materials.\cite{Datta_PRL97} It was in the study of the spectroscopy of semiconductors where it was also noticed a drawback of this technique, as the band gap was showed to be often misestimated  due to band bending effects present in these materials\cite{Stroscio_PRL86,Feenstra_09} and different normalization procedures\cite{Stroscio_PRL86, Koslowski_PRB07,Wagner_PRB07,Ziegler_PRB09} have to be used in order to obtain useful information.

A general procedure in STS measurements on molecular films is to average curves obtained over large areas in order to minimize effects on the LDOS due to local differences in morphology.\cite{Alvarado_PRL98} As a result, in most of the studies a representative I-V curve  is obtained, which gives information on the gap structure or on the different molecular levels of the molecular assembly, while the deviations from this value, that can be observed from the individual spectroscopic curves, are neglected. 
However these deviations, which are reflected in other spectroscopic techniques as in the case of the Photoluminescence (PL) characteristics, can give us information on the distribution of the LDOS, which deviates from its ideal one due to the local environment, impurities or defects.\cite{Vragovic_ChemPhys06}
In this letter we address this issue by comparing the Gap distribution obtained  by STS on a molecular material to its PL spectrum.   

In order to make our study we have chosen the extensively studied organic molecule N,N'-diphenyl-N,N'-bis(3-methylphenyl)-(1,1'-biphenyl)-4,4'-diamine, or TPD, which is a prototypical organic compound used in multilayer emitting devices as a hole transporting material.\cite{Bulovic_Nature96,Vragovic_ChemPhys06}
An interesting characteristic of this compound is its large Stokes shift of about 0.5 eV, what gives TPD a high transparency to its PL making it a good candidate for laser applications.\cite{Diaz_APL02} 
On the other hand, the intermolecular distances in its crystalline phase are rather large\cite{Kennedy_JMC02} and therefore the intermolecular interactions should play a minor role for the photophysics\cite{Scholz_JPC09} 

For our experiments, TPD films were evaporated simultaneously on 2.5x2.5 cm$^2$ fused silica substrates and on flame-annealed gold (111) deposited on glass. The fused silica substrates were used to obtain the  absorption and PL spectra of the compound. Absorption was measured in a Jasco V-650 spectrophotometer and the PL in a Jasco FP-6500/6600 fluorimeter, with the samples excited at 355 nm (3.49 eV) i.e. at the maximum energy of the lowest absorption band. 
For the electronic characterization of the sample surface we have used a homemade STM, built in the LT-Nanolab  at the University of Alicante with a PtIr tip, controlled with a Dulcinea Unit from Nanotec and the WSxM 5.0. \cite{Nanotec07} All experiments were done under ambient conditions. For the analysis of the curves, including Gap value and Fermi Energy, we have used the WSxM 5.0 and the shared-free program HiTim\cite{Nanolab}.

\begin{figure}[ht]
\epsfig{width=8cm,figure=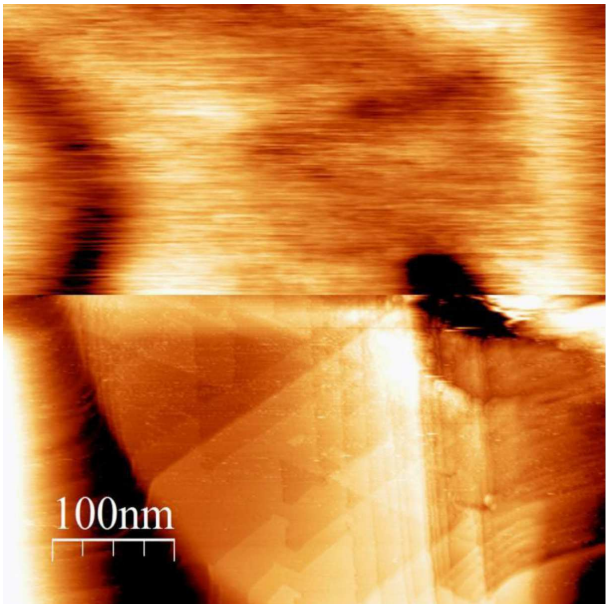}
\caption{\label{fig:STMimage} Typical STM image on the TPD films using a tunneling current of 10$^{-10}$A. The upper half of the image has been acquired using a positive Bias Voltage to the tip of 3V while for the lower part 100mV have been used. In the low-bias case the Au (111) substrate is imaged and the atomic terraces of gold are clearly seen since on these conditions TPD is non conductive.}
\end{figure}

An example of the obtained STM images is shown in Figure \ref{fig:STMimage}. These were performed on the TPD films using a positive Bias Voltage at the tip of about 3V above the expected energy gap of the TPD, and low currents ($\sim 1-5\cdot 10^{-10}$A). It has to be noticed that when a small Bias Voltage was used ($\sim 100$mV) we could resolve, in most of the cases, the Au (111) surface as there the TPD was not conducting (Shown in the lower part of Fig.~\ref{fig:STMimage} ).  The STM images show large areas of an homogenous flat film with hight differences of no more than 2 nm, separated by a sort of grain boundaries and other topographic details coming from the substrate topography. In general, the TPD film smoothness the roughness of the substrate. On this conditions I-V curves can be taken at different randomly distributed spots on the sample. We have performed these measurements over three different samples showing similar results.

\begin{figure}[ht]
\raggedright
\Large
 a)
\epsfig{width=7.2cm,figure=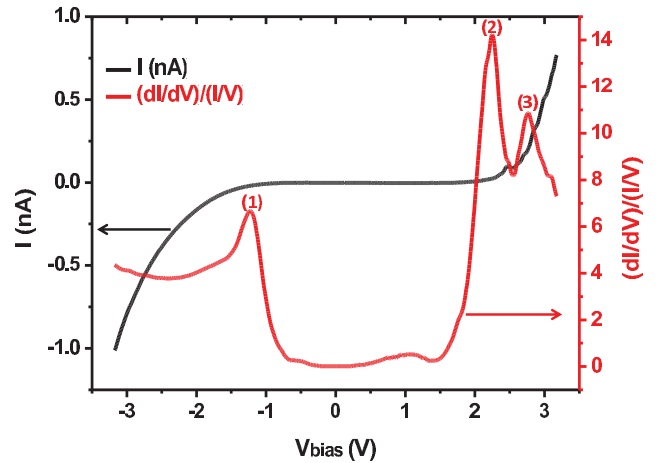, angle=-90}
b)

\epsfig{width=7.2cm,figure=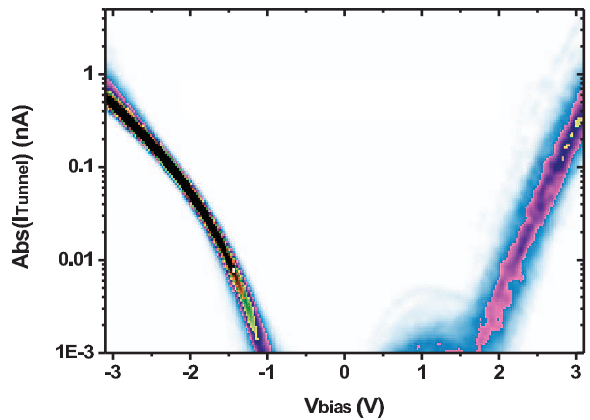, angle=-90 }
\caption{\label{fig:STScurves} Electronic transport characterization of the sample. Panel a) shows a typical IV curve taken with the STM starting at a current of 0.7 nA and its normalized derivative where  three main peaks characterizing the electronic properties of TPD are labeled. Panel b) shows in logarithmic scale a density plot of over two hundred IV curves taken at different spots of one of the samples. It can be noticed a difference in the distribution of the curves for positive and negative bias voltages.}
\end{figure}

The upper panel of Figure \ref{fig:STScurves} shows a typical I-V curve as obtained from our STM. It is possible to get information on the LDOS from the derivative of the I-V curve. In our case we used the normalization procedure by Stroscio {\it et al.}, \cite{Stroscio_PRL86} where the differential conductance is divided by the conductance $(dI/dV)/(I/V)$. It is important to stress here that special care had to be done in order to reduce the zero current point and the electrical noise to a level that would not influence the position of the spectroscopic peaks.  In our case, by this simple normalization procedure, three strong peaks are obtained which are labeled 1-3 in the upper panel of Fig.~\ref{fig:STScurves}. In order to minimize band-bending effects we have tried to perform the curves with the tip as far from the sample as possible, by using low tunnel currents of about $\sim 5\cdot 10^{-10}$ A. We have also checked the evolution of these peaks, as we increased the tunnel current for more than one order of magnitude, finding no significant variations ($<2\%$). 

\begin{figure}[ht]
\epsfig{width=7.5cm,figure=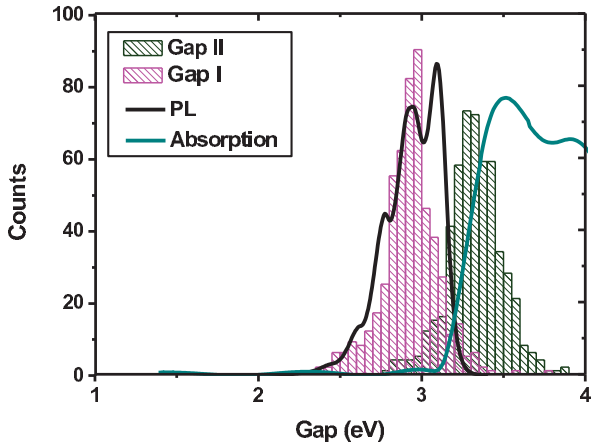, angle=-90}
\caption{\label{fig:STSdistribution} Distribution of the gap obtained from the analysis of 530 IV curves and the absorption and PL spectra of the sample.}
\end{figure}

The three peaks observed in the normalized derivative in Fig.~\ref{fig:STScurves}.a  correspond to the maximum of the LDOS coming from the molecular bands of the TPD. The first peak (1) is related to the top of the valence band related to the HOMO levels, being the closest to the Fermi energy thus confirming the n character of the semiconductor. The other two peaks, at positive Voltage values, are related to two electronic levels at the LUMO (2 and 3) which could correspond to the bottom of the conduction band of the ground(3) and excited (2) states of the molecules.  The levels above define the two possible electronic gaps of the molecules. In this way, the mean distance between peaks (1) and (2) is of 2.9 $\pm$ 0.2 eV  and between peaks (1) and (3) of 3.3 $\pm$ 0.2 eV defining respectively the main emitting and absorption peaks in the PL curve of the TPD (See continuos line on Fig.~\ref{fig:STSdistribution}). 

The "characteristic curve" of the TPD described above, however, shows slight shifts when taken at different spots of the sample. In Fig.~\ref{fig:STScurves}.b we have plotted, for the case of one of the samples,  over two hundred I-V curves together in a density plot with the current in logarithmic scale to visualize the emission gap. A dispersion of the values at which the electrical current becomes negligible is clearly seen around the average value which define the average gap (in dark). In this case, we already can see in this plot that most of the uncertainty comes from the position of the LUMO levels which show a Gaussian distribution centered at 1.9 $\pm$ 0.2 eV,\footnote{Error is defined by the half width of the Gaussian distribution.} while the HOMO is pined around 1.15 $\pm$ 0.09 eV, below the Fermi energy. This difference in the uncertainty of the two levels may come from the fact that the HOMO level is mostly localized on the central part of the molecule and is thus hardly sensitive to ring twists on the periphery of the molecule.\cite{Cornil_JPC01} Furthermore, the strongest dependence of the LUMO levels to the dihedral angle of the molecule\cite{Scholz_JPC09}  could make these more sensitive to its environment.

In order to quantify the distribution of the gap through the sample, we have made a histogram out of the values obtained from the I-V curves as shown in Figure \ref{fig:STSdistribution}. Remarkably, the measured electronic gap correspond to the PL curve of the sample with about the same uncertainty. Moreover, when the gap defined by the HOMO and the second LUMO peak is plotted, its histogram reassembles the absorption curve  of the molecular material. 

The slight shift of the gap distribution with respect the PL (less than a 5\%) may come from small deviations in the determination of the gap, coming form band bending effects. It is also very interesting to notice that occasionally we noticed a proportion of I-V curves giving band gap values well above the PL peak. This may be related to areas of the surface where the TPD molecules are not able to change their conformation and are not contributing to the Photoanalisis.
In this aspect, the STM analysis give us some extra information on the molecules, not given by other techniques and here the study of the whole gap distribution is essential to get this kind of information.
The results above were reproduced for the three samples and STM-tips used in our studies. However, we could notice slight variations in the position of the HOMO and LUMO peaks or in their distributions but not in the gap defined by these. These variations  could be attributed to differences in the tip that should be further studied. 
All the above shows that the STM characterization of the molecular film is not only telling which are the main electronic levels responsible, in this case, for the PL and absorption characteristics, but one also can learn from its distribution: first, about the levels which are most affected by the environment or configuration of the molecules and secondly, about the influence on the width of the luminescent characteristics of the material.   

In conclusion, we have shown that the characterization of the gap distribution, neglected up to now, of an organic semiconductor by STM can provide useful information that can help understanding the results provided by other physical measurements,  such as PL. In the case of our analysis of TPD films, we have fully characterized its electronic density of states around the gap by using a STM. The position with respect the Fermi energy of the valence and conduction bands coming from the HOMO and LUMO levels of the molecules have been studied through our samples and the levels responsible for the PL and absorption have been identified. We show that there is a source of uncertainty of the gap which comes mainly from a distribution in the LUMO levels of the molecular semiconductor, while the HOMO levels are shown to get a more defined value. This new source of uncertainty can complement the ones reported by other authors,\cite{Scholz_JPC09} providing a more complete picture of this effect and the possibility of better controlling it.

\begin{acknowledgments}
We thank A. Bensouici for instructive discussions on the characterization of semi conductive materials. We acknowledge financial support of the Spanish government (grants No. FIS2010-21883-C02-01, MAT2011-28167-C02-01 and No. CONSOLIDER CSD2007-0010) and Comunidad Valenciana
(ACOMP/2012/127 and PROMETEO/2012/011). This research has been partial funded by FEDER funds of the EU. 
\end{acknowledgments}

\bibliography{photo_stm}

\providecommand{\noopsort}[1]{}\providecommand{\singleletter}[1]{#1}%
\begin{thebibliography}{19}%
\makeatletter
\providecommand \@ifxundefined [1]{%
 \@ifx{#1\undefined}
}%
\providecommand \@ifnum [1]{%
 \ifnum #1\expandafter \@firstoftwo
 \else \expandafter \@secondoftwo
 \fi
}%
\providecommand \@ifx [1]{%
 \ifx #1\expandafter \@firstoftwo
 \else \expandafter \@secondoftwo
 \fi
}%
\providecommand \natexlab [1]{#1}%
\providecommand \enquote  [1]{``#1''}%
\providecommand \bibnamefont  [1]{#1}%
\providecommand \bibfnamefont [1]{#1}%
\providecommand \citenamefont [1]{#1}%
\providecommand \href@noop [0]{\@secondoftwo}%
\providecommand \href [0]{\begingroup \@sanitize@url \@href}%
\providecommand \@href[1]{\@@startlink{#1}\@@href}%
\providecommand \@@href[1]{\endgroup#1\@@endlink}%
\providecommand \@sanitize@url [0]{\catcode `\\12\catcode `\$12\catcode
  `\&12\catcode `\#12\catcode `\^12\catcode `\_12\catcode `\%12\relax}%
\providecommand \@@startlink[1]{}%
\providecommand \@@endlink[0]{}%
\providecommand \url  [0]{\begingroup\@sanitize@url \@url }%
\providecommand \@url [1]{\endgroup\@href {#1}{\urlprefix }}%
\providecommand \urlprefix  [0]{URL }%
\providecommand \Eprint [0]{\href }%
\providecommand \doibase [0]{http://dx.doi.org/}%
\providecommand \selectlanguage [0]{\@gobble}%
\providecommand \bibinfo  [0]{\@secondoftwo}%
\providecommand \bibfield  [0]{\@secondoftwo}%
\providecommand \translation [1]{[#1]}%
\providecommand \BibitemOpen [0]{}%
\providecommand \bibitemStop [0]{}%
\providecommand \bibitemNoStop [0]{.\EOS\space}%
\providecommand \EOS [0]{\spacefactor3000\relax}%
\providecommand \BibitemShut  [1]{\csname bibitem#1\endcsname}%
\let\auto@bib@innerbib\@empty
\bibitem [{\citenamefont {Binning}\ \emph
  {et~al.}(1982{\natexlab{a}})\citenamefont {Binning}, \citenamefont {Rohrer},
  \citenamefont {Gerber},\ and\ \citenamefont {Weibel}}]{Binning_82PRL}%
  \BibitemOpen
  \bibfield  {author} {\bibinfo {author} {\bibfnamefont {G.}~\bibnamefont
  {Binning}}, \bibinfo {author} {\bibfnamefont {H.}~\bibnamefont {Rohrer}},
  \bibinfo {author} {\bibfnamefont {{\relax Ch}.}~\bibnamefont {Gerber}}, \
  and\ \bibinfo {author} {\bibfnamefont {E.}~\bibnamefont {Weibel}},\
  }\href@noop {} {\bibfield  {journal} {\bibinfo  {journal} {Phys.\ Rev.\
  Lett.}\ }\textbf {\bibinfo {volume} {49}},\ \bibinfo {pages} {57} (\bibinfo
  {year} {1982}{\natexlab{a}})}\BibitemShut {NoStop}%
\bibitem [{\citenamefont {Binning}\ \emph
  {et~al.}(1982{\natexlab{b}})\citenamefont {Binning}, \citenamefont {Rohrer},
  \citenamefont {Gerber},\ and\ \citenamefont {Weibel}}]{Binning_82JAP}%
  \BibitemOpen
  \bibfield  {author} {\bibinfo {author} {\bibfnamefont {G.}~\bibnamefont
  {Binning}}, \bibinfo {author} {\bibfnamefont {H.}~\bibnamefont {Rohrer}},
  \bibinfo {author} {\bibfnamefont {{\relax Ch}.}~\bibnamefont {Gerber}}, \
  and\ \bibinfo {author} {\bibfnamefont {E.}~\bibnamefont {Weibel}},\
  }\href@noop {} {\bibfield  {journal} {\bibinfo  {journal} {J.\ Appl.\ Phys.}\
  }\textbf {\bibinfo {volume} {40}},\ \bibinfo {pages} {178} (\bibinfo {year}
  {1982}{\natexlab{b}})}\BibitemShut {NoStop}%
\bibitem [{\citenamefont {Stroscio}, \citenamefont {Feenstra},\ and\
  \citenamefont {Fein}(1986)}]{Stroscio_PRL86}%
  \BibitemOpen
  \bibfield  {author} {\bibinfo {author} {\bibfnamefont {J.~A.}\ \bibnamefont
  {Stroscio}}, \bibinfo {author} {\bibfnamefont {R.~M.}\ \bibnamefont
  {Feenstra}}, \ and\ \bibinfo {author} {\bibfnamefont {A.}~\bibnamefont
  {Fein}},\ }\href@noop {} {\bibfield  {journal} {\bibinfo  {journal} {Phys.\
  Rev.\ Lett}\ }\textbf {\bibinfo {volume} {57}},\ \bibinfo {pages} {2579}
  (\bibinfo {year} {1986})}\BibitemShut {NoStop}%
\bibitem [{\citenamefont {Feenstra}(1994)}]{Feenstra_94}%
  \BibitemOpen
  \bibfield  {author} {\bibinfo {author} {\bibfnamefont {R.~M.}\ \bibnamefont
  {Feenstra}},\ }\href@noop {} {\bibfield  {journal} {\bibinfo  {journal}
  {Surf.\ Sci.}\ }\textbf {\bibinfo {volume} {299/300}},\ \bibinfo {pages}
  {965} (\bibinfo {year} {1994})}\BibitemShut {NoStop}%
\bibitem [{\citenamefont {Datta}\ \emph {et~al.}(1997)\citenamefont {Datta},
  \citenamefont {Tian}, \citenamefont {Hong}, \citenamefont {Reifenberger},
  \citenamefont {Henderson},\ and\ \citenamefont {Kubiak}}]{Datta_PRL97}%
  \BibitemOpen
  \bibfield  {author} {\bibinfo {author} {\bibfnamefont {S.}~\bibnamefont
  {Datta}}, \bibinfo {author} {\bibfnamefont {W.}~\bibnamefont {Tian}},
  \bibinfo {author} {\bibfnamefont {S.}~\bibnamefont {Hong}}, \bibinfo {author}
  {\bibfnamefont {R.}~\bibnamefont {Reifenberger}}, \bibinfo {author}
  {\bibfnamefont {J.~I.}\ \bibnamefont {Henderson}}, \ and\ \bibinfo {author}
  {\bibfnamefont {C.~P.}\ \bibnamefont {Kubiak}},\ }\href@noop {} {\bibfield
  {journal} {\bibinfo  {journal} {Phys.\ Rev.\ Lett.}\ }\textbf {\bibinfo
  {volume} {79}},\ \bibinfo {pages} {2530} (\bibinfo {year}
  {1997})}\BibitemShut {NoStop}%
\bibitem [{\citenamefont {Feenstra}(2009)}]{Feenstra_09}%
  \BibitemOpen
  \bibfield  {author} {\bibinfo {author} {\bibfnamefont {R.~M.}\ \bibnamefont
  {Feenstra}},\ }\href@noop {} {\bibfield  {journal} {\bibinfo  {journal}
  {Surf.\ Sci.}\ }\textbf {\bibinfo {volume} {603}},\ \bibinfo {pages} {2841}
  (\bibinfo {year} {2009})}\BibitemShut {NoStop}%
\bibitem [{\citenamefont {Koslowski}\ \emph {et~al.}(2007)\citenamefont
  {Koslowski}, \citenamefont {Dietrich}, \citenamefont {Tschetschetkin},\ and\
  \citenamefont {Ziemann}}]{Koslowski_PRB07}%
  \BibitemOpen
  \bibfield  {author} {\bibinfo {author} {\bibfnamefont {B.}~\bibnamefont
  {Koslowski}}, \bibinfo {author} {\bibfnamefont {C.}~\bibnamefont {Dietrich}},
  \bibinfo {author} {\bibfnamefont {A.}~\bibnamefont {Tschetschetkin}}, \ and\
  \bibinfo {author} {\bibfnamefont {P.}~\bibnamefont {Ziemann}},\ }\href
  {\doibase 10.1103/PhysRevB.75.035421} {\bibfield  {journal} {\bibinfo
  {journal} {Phys. Rev. B}\ }\textbf {\bibinfo {volume} {75}},\ \bibinfo
  {pages} {035421} (\bibinfo {year} {2007})}\BibitemShut {NoStop}%
\bibitem [{\citenamefont {Wagner}, \citenamefont {Franke},\ and\ \citenamefont
  {Fritz}(2007)}]{Wagner_PRB07}%
  \BibitemOpen
  \bibfield  {author} {\bibinfo {author} {\bibfnamefont {C.}~\bibnamefont
  {Wagner}}, \bibinfo {author} {\bibfnamefont {R.}~\bibnamefont {Franke}}, \
  and\ \bibinfo {author} {\bibfnamefont {T.}~\bibnamefont {Fritz}},\ }\href
  {\doibase 10.1103/PhysRevB.75.235432} {\bibfield  {journal} {\bibinfo
  {journal} {Phys. Rev. B}\ }\textbf {\bibinfo {volume} {75}},\ \bibinfo
  {pages} {235432} (\bibinfo {year} {2007})}\BibitemShut {NoStop}%
\bibitem [{\citenamefont {Ziegler}\ \emph {et~al.}(2009)\citenamefont
  {Ziegler}, \citenamefont {N\'eel}, \citenamefont {Sperl}, \citenamefont
  {Kr\"oger},\ and\ \citenamefont {Berndt}}]{Ziegler_PRB09}%
  \BibitemOpen
  \bibfield  {author} {\bibinfo {author} {\bibfnamefont {M.}~\bibnamefont
  {Ziegler}}, \bibinfo {author} {\bibfnamefont {N.}~\bibnamefont {N\'eel}},
  \bibinfo {author} {\bibfnamefont {A.}~\bibnamefont {Sperl}}, \bibinfo
  {author} {\bibfnamefont {J.}~\bibnamefont {Kr\"oger}}, \ and\ \bibinfo
  {author} {\bibfnamefont {R.}~\bibnamefont {Berndt}},\ }\href {\doibase
  10.1103/PhysRevB.80.125402} {\bibfield  {journal} {\bibinfo  {journal} {Phys.
  Rev. B}\ }\textbf {\bibinfo {volume} {80}},\ \bibinfo {pages} {125402}
  (\bibinfo {year} {2009})}\BibitemShut {NoStop}%
\bibitem [{\citenamefont {Alvarado}\ \emph {et~al.}(1998)\citenamefont
  {Alvarado}, \citenamefont {Seidler}, \citenamefont {Lidzey},\ and\
  \citenamefont {Bradley}}]{Alvarado_PRL98}%
  \BibitemOpen
  \bibfield  {author} {\bibinfo {author} {\bibfnamefont {S.~F.}\ \bibnamefont
  {Alvarado}}, \bibinfo {author} {\bibfnamefont {P.}~\bibnamefont {Seidler}},
  \bibinfo {author} {\bibfnamefont {D.}~\bibnamefont {Lidzey}}, \ and\ \bibinfo
  {author} {\bibfnamefont {D.~D.~C.}\ \bibnamefont {Bradley}},\ }\href@noop {}
  {\bibfield  {journal} {\bibinfo  {journal} {Phys.\ Rev.\ Lett.}\ }\textbf
  {\bibinfo {volume} {81}},\ \bibinfo {pages} {1082} (\bibinfo {year}
  {1998})}\BibitemShut {NoStop}%
\bibitem [{\citenamefont {Vragoci\'c}, \citenamefont {Calzado},\ and\
  \citenamefont {D\'{\i}az-Garc\'{\i}a}(2006)}]{Vragovic_ChemPhys06}%
  \BibitemOpen
  \bibfield  {author} {\bibinfo {author} {\bibfnamefont {I.}~\bibnamefont
  {Vragoci\'c}}, \bibinfo {author} {\bibfnamefont {E.~M.}\ \bibnamefont
  {Calzado}}, \ and\ \bibinfo {author} {\bibfnamefont {M.~A.}\ \bibnamefont
  {D\'{\i}az-Garc\'{\i}a}},\ }\href@noop {} {\bibfield  {journal} {\bibinfo
  {journal} {Chem.\ Phys.}\ }\textbf {\bibinfo {volume} {332}},\ \bibinfo
  {pages} {48} (\bibinfo {year} {2006})}\BibitemShut {NoStop}%
\bibitem [{\citenamefont {Bulovi\'c}\ \emph {et~al.}(1996)\citenamefont
  {Bulovi\'c}, \citenamefont {Gu}, \citenamefont {Burrows}, \citenamefont
  {Forest},\ and\ \citenamefont {Thompson}}]{Bulovic_Nature96}%
  \BibitemOpen
  \bibfield  {author} {\bibinfo {author} {\bibfnamefont {V.}~\bibnamefont
  {Bulovi\'c}}, \bibinfo {author} {\bibfnamefont {G.}~\bibnamefont {Gu}},
  \bibinfo {author} {\bibfnamefont {P.~E.}\ \bibnamefont {Burrows}}, \bibinfo
  {author} {\bibfnamefont {S.~R.}\ \bibnamefont {Forest}}, \ and\ \bibinfo
  {author} {\bibfnamefont {M.~E.}\ \bibnamefont {Thompson}},\ }\href@noop {}
  {\bibfield  {journal} {\bibinfo  {journal} {Nature}\ }\textbf {\bibinfo
  {volume} {380}},\ \bibinfo {pages} {29} (\bibinfo {year} {1996})}\BibitemShut
  {NoStop}%
\bibitem [{\citenamefont {D\'{\i}az-Garc\'{\i}a}, \citenamefont {de~\'Avila},\
  and\ \citenamefont {Kuzyk}(2002)}]{Diaz_APL02}%
  \BibitemOpen
  \bibfield  {author} {\bibinfo {author} {\bibfnamefont {M.~A.}\ \bibnamefont
  {D\'{\i}az-Garc\'{\i}a}}, \bibinfo {author} {\bibfnamefont {S.~F.}\
  \bibnamefont {de~\'Avila}}, \ and\ \bibinfo {author} {\bibfnamefont {M.~G.}\
  \bibnamefont {Kuzyk}},\ }\href@noop {} {\bibfield  {journal} {\bibinfo
  {journal} {Appl. Phys. Lett.}\ }\textbf {\bibinfo {volume} {80}},\ \bibinfo
  {pages} {4486} (\bibinfo {year} {2002})}\BibitemShut {NoStop}%
\bibitem [{\citenamefont {Kennedy}\ \emph {et~al.}(2002)\citenamefont
  {Kennedy}, \citenamefont {Smith}, \citenamefont {Tackley}, \citenamefont
  {David}, \citenamefont {Shankland}, \citenamefont {Brown},\ and\
  \citenamefont {Teat}}]{Kennedy_JMC02}%
  \BibitemOpen
  \bibfield  {author} {\bibinfo {author} {\bibfnamefont {A.~R.}\ \bibnamefont
  {Kennedy}}, \bibinfo {author} {\bibfnamefont {W.~E.}\ \bibnamefont {Smith}},
  \bibinfo {author} {\bibfnamefont {D.~R.}\ \bibnamefont {Tackley}}, \bibinfo
  {author} {\bibfnamefont {W.~I.~F.}\ \bibnamefont {David}}, \bibinfo {author}
  {\bibfnamefont {K.}~\bibnamefont {Shankland}}, \bibinfo {author}
  {\bibfnamefont {B.}~\bibnamefont {Brown}}, \ and\ \bibinfo {author}
  {\bibfnamefont {S.~J.}\ \bibnamefont {Teat}},\ }\href@noop {} {\bibfield
  {journal} {\bibinfo  {journal} {J. Matter. Chem.}\ }\textbf {\bibinfo
  {volume} {12}},\ \bibinfo {pages} {168} (\bibinfo {year} {2002})}\BibitemShut
  {NoStop}%
\bibitem [{\citenamefont {Scholz}\ \emph {et~al.}(2009)\citenamefont {Scholz},
  \citenamefont {Gissl\'en}, \citenamefont {Himcinschi}, \citenamefont
  {Vragoci\'c}, \citenamefont {Calzado}, \citenamefont {Louis}, \citenamefont
  {{San Fabi\'an Maroto}},\ and\ \citenamefont
  {D\'{\i}az-Garc\'{\i}a}}]{Scholz_JPC09}%
  \BibitemOpen
  \bibfield  {author} {\bibinfo {author} {\bibfnamefont {R.}~\bibnamefont
  {Scholz}}, \bibinfo {author} {\bibfnamefont {L.}~\bibnamefont {Gissl\'en}},
  \bibinfo {author} {\bibfnamefont {C.}~\bibnamefont {Himcinschi}}, \bibinfo
  {author} {\bibfnamefont {I.}~\bibnamefont {Vragoci\'c}}, \bibinfo {author}
  {\bibfnamefont {E.~M.}\ \bibnamefont {Calzado}}, \bibinfo {author}
  {\bibfnamefont {E.}~\bibnamefont {Louis}}, \bibinfo {author} {\bibfnamefont
  {E.}~\bibnamefont {{San Fabi\'an Maroto}}}, \ and\ \bibinfo {author}
  {\bibfnamefont {M.~A.}\ \bibnamefont {D\'{\i}az-Garc\'{\i}a}},\ }\href@noop
  {} {\bibfield  {journal} {\bibinfo  {journal} {J.\ Phys.\ Chem.\ A}\ }\textbf
  {\bibinfo {volume} {113}},\ \bibinfo {pages} {315} (\bibinfo {year}
  {2009})}\BibitemShut {NoStop}%
\bibitem [{\citenamefont {Horcas}\ \emph {et~al.}(2007)\citenamefont {Horcas},
  \citenamefont {Fernandez}, \citenamefont {Gomez-Rodriguez}, \citenamefont
  {Colchero}, \citenamefont {Gomez-Herrero},\ and\ \citenamefont
  {Baro}}]{Nanotec07}%
  \BibitemOpen
  \bibfield  {author} {\bibinfo {author} {\bibfnamefont {I.}~\bibnamefont
  {Horcas}}, \bibinfo {author} {\bibfnamefont {R.}~\bibnamefont {Fernandez}},
  \bibinfo {author} {\bibfnamefont {J.}~\bibnamefont {Gomez-Rodriguez}},
  \bibinfo {author} {\bibfnamefont {J.}~\bibnamefont {Colchero}}, \bibinfo
  {author} {\bibfnamefont {J.}~\bibnamefont {Gomez-Herrero}}, \ and\ \bibinfo
  {author} {\bibfnamefont {A.~M.}\ \bibnamefont {Baro}},\ }\href@noop {}
  {\bibfield  {journal} {\bibinfo  {journal} {Rev.\ Sci.\ Instrum.}\ }\textbf
  {\bibinfo {volume} {78}},\ \bibinfo {pages} {013705} (\bibinfo {year}
  {2007})}\BibitemShut {NoStop}%
\bibitem [{\citenamefont {Untiedt}()}]{Nanolab}%
  \BibitemOpen
  \bibfield  {author} {\bibinfo {author} {\bibfnamefont {C.}~\bibnamefont
  {Untiedt}},\ }\href@noop {} {}\bibinfo {note} {Shared free HiTim program,
  found at www.ua.es/personal/untiedt}\BibitemShut {NoStop}%
\bibitem [{Note1()}]{Note1}%
  \BibitemOpen
  \bibinfo {note} {Error is defined by the half width of the Gaussian
  distribution.}\BibitemShut {Stop}%
\bibitem [{\citenamefont {Cornil}\ \emph {et~al.}(2001)\citenamefont {Cornil},
  \citenamefont {Gruhn}, \citenamefont {{dos Santos}}, \citenamefont
  {Malagoli}, \citenamefont {Lee}, \citenamefont {Barlow}, \citenamefont
  {Thayumanavan}, \citenamefont {Marder}, \citenamefont {Armstrong},\ and\
  \citenamefont {Br\'edas}}]{Cornil_JPC01}%
  \BibitemOpen
  \bibfield  {author} {\bibinfo {author} {\bibfnamefont {J.}~\bibnamefont
  {Cornil}}, \bibinfo {author} {\bibfnamefont {N.~E.}\ \bibnamefont {Gruhn}},
  \bibinfo {author} {\bibfnamefont {D.~A.}\ \bibnamefont {{dos Santos}}},
  \bibinfo {author} {\bibfnamefont {M.}~\bibnamefont {Malagoli}}, \bibinfo
  {author} {\bibfnamefont {P.~A.}\ \bibnamefont {Lee}}, \bibinfo {author}
  {\bibfnamefont {S.}~\bibnamefont {Barlow}}, \bibinfo {author} {\bibfnamefont
  {S.}~\bibnamefont {Thayumanavan}}, \bibinfo {author} {\bibfnamefont {S.~R.}\
  \bibnamefont {Marder}}, \bibinfo {author} {\bibfnamefont {N.~R.}\
  \bibnamefont {Armstrong}}, \ and\ \bibinfo {author} {\bibfnamefont {J.~L.}\
  \bibnamefont {Br\'edas}},\ }\href@noop {} {\bibfield  {journal} {\bibinfo
  {journal} {J. Phys. Chem. A}\ }\textbf {\bibinfo {volume} {105}},\ \bibinfo
  {pages} {5206} (\bibinfo {year} {2001})}\BibitemShut {NoStop}%
\end{thebibliography}%

\end{document}